\documentstyle[aps,prl,twocolumn,floats,graphicx]{revtex}

\begin{document}
\draft
\preprint{}
\wideabs{
\title{Nodal Quasiparticle Lifetime in the Superconducting State of ${\bf Bi_{2}Sr_{2}CaCu_{2}O_{8+\delta}}$}
\author{J. Corson, J. Orenstein}
\address{Materials Sciences Division, Lawrence Berkeley National Laboratory and Physics Department, University
of California, Berkeley, California 94720}
\author{Seongshik Oh, J. O'Donnell, J. N. Eckstein}
\address{Department of Physics, University of Illinois, Urbana, Illinois 61801}
\date{\today}
\maketitle
\begin{abstract}
We have measured the complex conductivity, $\sigma$, of a $Bi_{2}Sr_{2}CaCu_{2}O_{8+\delta}$ (BSCCO) thin film between 0.2 and 0.8 THz.  We find $\sigma$ in the superconducting state to be well described as the sum of contributions from quasiparticles, the condensate, and order parameter fluctuations which draw 30$\%$ of the spectral weight from the condensate.  An analysis based on this decomposition yields a quasiparticle scattering rate on the order of $k_{B}T/\hbar$ for temperatures below $T_c$.\end{abstract}

\pacs{74.25.Gz, 74.25.Nf, 74.25.-q, 74.72.Hs}
}

The unusual properties of the quasiparticle (QP) lifetime $\tau_{qp}$ in high-$T_{c}$ cuprate superconductors provide evidence for a non-Fermi liquid normal state and a non-BCS mechanism.  Because angle-resolved photoemission (ARPES) shows that QP properties are highly anisotropic\cite{Shen,Randeria}, it is important to distinguish the behavior of $\tau_{qp}$ at different parts of the Fermi surface.  Near the maximum of the d-wave gap at the ($\pi$, 0) point in momentum space, the ARPES lineshape narrows rapidly below $T_{c}$\cite{Shen,Randeria,Federov}.  The change suggests the onset of a well-defined QP in this antinodal region.  Recently, improvements in detectors have made it possible to resolve the lineshape in the nodal or ($\pi,\pi$) direction as well.  Valla {\em et al.}\cite{Valla} reported that in the normal state, $\tau_{qp}$ can be described by the marginal Fermi liquid phenomenology\cite{Varma}, $1/\tau_{qp}\sim$ max$(\omega, T)$.  Surprisingly, $1/\tau_{qp}$ of nodal quasiparticles (at the Fermi surface) remains approximately proportional to $T$ in the superconducting state as well\cite{Valla,Kaminski}.  Thus the change in the lifetime of such QP's upon entering the superconducting state is much less than that of the antinodal QPs.  

The nodal QP's apparent insensitivity to superconductivity is surprising for two reasons.  First, one might expect the reduction in the phase space for scattering which accompanies the formation of a gap to increase the QP lifetime.  Second, $\tau_{qp}$, as determined by microwave \cite{Nuss,Bonn,Hosseini} and thermal conductivity \cite{Thermal} measurements in the closely related compound $YBa_{2}Cu_{3}O_{7-\delta}$ (YBCO) does indeed increase rapidly below $T_{c}$.  While such transport experiments are not momentum resolved, there is strong evidence that they probe the nodal QP lifetime.  It was reported in Ref. \cite{Hosseini} that the microwave conductivity, $\sigma=\sigma_{1}+i\sigma_{2}$, can be described (at low $T$) by a two-fluid model which sums the contributions of condensate and QPs.  The spectral weight in the normal fluid, $\rho_{n}\sim\int\sigma_{1}d\nu$, grows linearly with $T$ at low temperature at a rate which matches the linear decrease in superfluid density $\rho_{s}(T)$, measured independently via the penetration depth.  The linear rate of transfer from $\rho_{s}$ to $\rho_{n}$ agrees with the theoretical prediction for nodal QPs in a d-wave superconductor\cite{DWaveTheory}.

Thus, the transport measurements of $\tau_{qp}$ in YBCO stand in apparent contradiction to ARPES measurements in BSCCO.  It would appear that either the single-particle lifetime probed by ARPES is qualitatively different than the transport lifetime, or that $\tau_{qp}$ differs in BSCCO and YBCO despite their similar bilayer structure and $T_{c}$. The resolution of this issue requires interchanging the two probes and the two material systems.  Since ARPES lineshapes in YBCO are difficult to obtain, it is important to focus on the transport properties of BSCCO, for example the microwave conductivity\cite{Trunin}.

Unfortunately, it has proved difficult to determine $\tau_{qp}$ reliably from microwave data in BSCCO.  In contrast to the strong dispersion in YBCO, $\sigma_1$ is approximately constant in the range from $\nu$=10-50 GHz\cite{Sridhar,Morgan,Maeda}.  Consequently, the transport lifetime is not directly resolved in the conductivity spectra, although they do indicate that $\tau_{qp}\ll 1/2\pi\cdot$50 GHz, which is much smaller than in YBCO.  Despite the lack of frequency dependence, in principle $\tau_{qp}$ could still be estimated from the magnitude of $\sigma_1$, which (in the $\omega\tau_{qp}\ll1$ limit) is proportional to $\rho_{n}\tau_{qp}$.  Using the two-fluid model, the magnitude of $\rho_n (T)$ follows from measurement of $\rho_s (T)$.  In practice, this analysis yields results which are not self-consistent, for the following reason.  Just as in YBCO, the condensate density varies as $\rho_s(0)-\alpha T$ at low T\cite{Sridhar,Morgan,Maeda}, implying that the samples are in the clean d-wave regime where $\rho_n=\alpha T$.  However, $\sigma_1$ in BSCCO does not tend to zero as $T$ approaches zero. Even at the lowest temperatures measured, about 5 K, $\sigma_1$ remains approximately 8 times larger than the normal state value just above $T_c$, $\sigma_n (T_c)$, and is far larger than the `universal conductivity' proposed by Lee\cite{Lee}. Thus $\tau_{qp}$ calculated from $\sigma_1/\rho_{n}$ appears to diverge, which is inconsistent with the observation that the conductivity is frequency independent in the microwave regime.  To extract a finite $\tau_{qp}$ one may assume that a large fraction of the quasiparticles remain uncondensed, i.e. $\rho_n(T)=\rho_n(0)+\alpha T$\cite{Maeda}, or subtract the zero temperature limit of the dissipation prior to the two-fluid model analysis\cite{Shibauchi}.

In view of the uncertainties inherent in these assumptions, it is desirable to extend the conductivity measurements to the frequency range probed by ARPES, $\approx k_{B}T/\hbar$.  Here we report measurements of $\sigma$ in a BSCCO thin film in the frequency range from 0.18 to 1.0 THz, above the domain of microwave and below that of infrared spectroscopies.  To cover this region, we used time-domain terahertz spectroscopy, a technique which is based on the generation and detection of single cycle electromagnetic pulses.  The transmission coefficient of such pulses yields both the real and imaginary parts of $\sigma$ directly and independently\cite{JoeTHzChapter}. The sample, with $T_{c}$=$85 K$, was grown on a $LaAlO_3$ substrate using atomic layer-by-layer molecular beam epitaxy\cite{JimSampleGrowth}. The resistance $R$ versus $T$ is linear and $R(0)/R(300K)=4\times 10^{-2}$, where $R(0)$ is the extrapolation of the linear resistance to zero temperature.

The terahertz data show a strong frequency dependence in $\sigma$ below $T_{c}$, from which $\tau_{qp}$ in the superconducting state can be determined directly.  \begin{figure}[h]
     \includegraphics[width=3.25in]{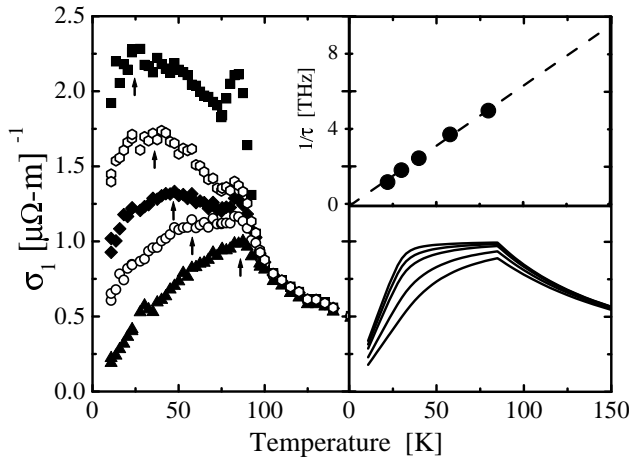}
\caption{Left panel: $\sigma_{1}$ plotted versus $T$ for 0.2, 0.3, 0.4, 0.6 and 0.8 THz as squares, octagons, diamonds, circles and triangles.  At each frequency the arrows mark the $T$ where $\sigma_{1}$ begins to decrease.  Upper right panel: $1/\tau_{qp}$ implied by the arrows plotted versus $T$, with the same $T$ scale as the panel below.  The dashed line is a fit to $1/\tau_{qp}\propto T$.  Lower right panel: Drude conductivity, $\sigma_{qp}(T)$, using $\tau_{qp}$ proportional to the linear fit in the upper panel plotted on the same scale as the left panel. The same frequencies are shown as the left panel, increasing $\nu$ having smaller $\sigma_{qp}(T)$.}
\label{fig:First}
\end{figure}We find that $1/\tau_{qp}(T)\approx 0.8k_{B}(T+10K)/\hbar$, consistent with ARPES measurements \cite{Valla,Kaminski}.  Furthermore, we find the total conductivity to be incompatible with the two-fluid model.  Instead, our measurements indicate that the conductivity comprises three contributions: QPs, the condensate, and a low frequency collective mode with spectral weight drawn from the condensate.

We now turn to $\sigma_{1}(\omega,T)$ and discuss its salient features.  The left panel of Figure 1 shows $\sigma_{1}$ plotted versus $T$ for a representative sample of frequencies in our range.  At temperatures above $T_{c}$ there is little frequency dependence, as we would expect since $1/\tau_{qp} \gg 2\pi\cdot$1 THz in this temperature range.  Below $T_{c}$, however, $\sigma_{1}(T)$ depends strongly on frequency.  At $\nu$=0.2 THz the $T$ dependence is quite similar to that measured in the microwave regime on BSCCO single crystals \cite{Sridhar,Morgan,Maeda}.  $\sigma_{1}$ rises to a maximum at 25 K, but at 5 K remains much greater than $\sigma_{n}(T_{c})$.  As the frequency increases from 0.2 to 0.8 THz the maximum shifts from 25 K to $T_{c}$, indicating that $1/\tau_{qp}$ is sweeping through our bandwidth.  For a given frequency, the QP conductivity will decrease below the temperature where $1/\tau_{qp}(T)\approx \omega$. At each measurement frequency, this $T$ is marked with an arrow.  The measurement frequency (in radians/s) and corresponding $T$ for each arrow are shown in the upper right panel of Fig. 1.  This admittedly rough estimate clearly suggests a QP lifetime with the form $1/\tau_{qp}\sim T$.  Finally, the peak in $\sigma_{1}$ near $T_{c}$, most visible at low frequencies, has been attributed to phase fluctuations of the order parameter due to thermal vortex, anti-vortex pairs\cite{Nature}. 

Following its success in YBCO, we attempt to describe the conductivity using a two fluid model, in which $\sigma$ below $T_{c}$ is comprised of condensate and QP contributions.  The condensate's real, or dissipative, conductivity is a $\delta$-function at $\nu=0$.  Therefore, for any nonzero frequency $\sigma_{1}$ is due solely to the QPs.  Following the analysis of YBCO data, we choose a Drude conductivity for the QPs, $\sigma_{qp}(\omega,T)/\sigma_{Q}\equiv(\rho_{n}\tau_{qp}/\hbar)/(1+\omega^{2}\tau_{qp}^{2})$.  Here $\sigma_{Q}\equiv e^{2}/(\hbar d)$ is the quantum conductivity of a stack of bilayers with spacing $d=15.4 \AA$.  The spectral weight, $\rho_{n}$ is expressed in units of energy as will be described below.    

The two $T$ dependent parameters, $\tau_{qp}(T)$ and $\rho_{n}(T)$ are clearly suggested by the data.  Regarding $\tau_{qp}(T)$, the upper right panel of Fig. 1 suggests that $1/\tau_{qp}\sim T$ over a wide temperature range.  
The form of $\rho_{n}(T)$ follows from the $T$ dependence of the superfluid density.  We obtain $\rho_{s}(T)$ from the imaginary part of the conductivity, $\sigma_{2}$, which is measured independently of $\sigma_{1}$ in our experiment.  We observe that $\rho_{s}(T)=\rho_{0}-\alpha T$ below about 30 K, suggesting that $\rho_{n}=\alpha T$ at low $T$.  For purposes of modeling, we have found that the most accurate description of the terahertz conductivity results if we assume that $\rho_{n}=\alpha T$ up to $T_{c}$.  Of course the total normal fluid weight must increase faster than this in order that $\rho_{n}+\rho_{s}$ remains constant as $T$ changes.  However, we are modeling only the spectral weight contained in a Lorentzian component centered at zero frequency. 
 
In the right-hand panel of Fig. 1, we plot $\sigma_{qp}(T)$, the Drude conductivity with $\rho_{n}(T)$ and $\tau_{qp}(T)$ as described above.  Focusing first on the highest frequencies in our range, we see that $\sigma_{qp}(T)$ agrees rather well with the observed $\sigma_{1}(T)$.  On the other hand, $\sigma_{1}$ at low frequency is much larger than predicted by a Drude model for the QP conductivity.  In fact, we have found it impossible to describe the conductivity in the range from 0.2-1 THz with a two-fluid model ($\rho_n(T)+\rho_s(T)$ is constant) regardless of the choice of $\tau_{qp}(T)$.  The difficulty is that the spectral weight in the low-frequency end of our range increases with decreasing $T$, rather than decreasing as expected for the normal fluid component in a two-fluid model.  The conductivity spectra suggest the presence of an additional component, whose spectral weight increases with decreasing $T$.  (Simulations based on \cite{Feenstra} show that the spectral weight at low frequency cannot be caused by inadvertent coupling to the c-axis plasmon due to off-normal incidence).      
  \begin{figure}[h]
     \includegraphics[width=3.25in]{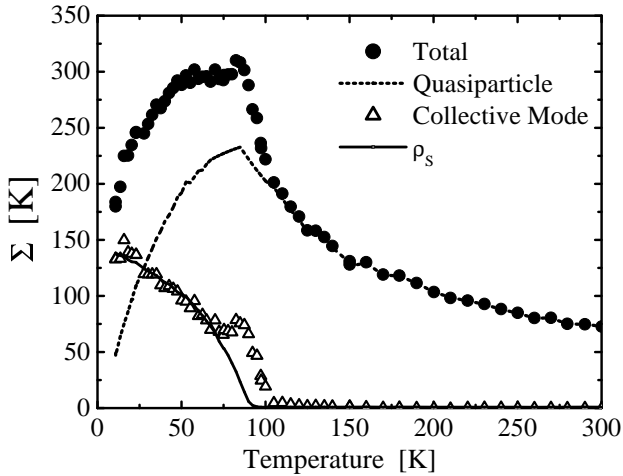}
\caption{The spectral weight of the total conductivity from 0.2-0.8 THz, $\Sigma(T)$, the QP Drude conductivity, $\Sigma_{qp}(T)$, and the difference between the two, $\Sigma(T)-\Sigma_{qp}(T)$.  The solid line shows the superfluid density, $\rho_{s}(T)$, multiplied by 0.18, showing the proportionality between this and the portion of the spectral weight not due to the QPs.}
\label{fig:Second}
\end{figure}

To help identify the additional component of $\sigma$, we consider the difference between the measured spectral weight and the amount ascribed to QPs using the parameters given above.  Fig. 2 shows the spectral weight measured in our bandwidth, $\Sigma(T)$, and that due to the normal fluid, $\Sigma_{qp}(T)$, plotted versus $T$.  Because of the extreme anisotropy of BSCCO we express spectral weight in terms of an areal density per CuO$_{2}$ bilayer.  The areal density is proportional to the phase or charge stiffness per bilayer, and can be conveniently expressed in thermal units of energy, or Kelvins.  The spectral weights denoted by $\Sigma$ are distinct from those discussed earlier labeled with $\rho$'s.  Whereas $\rho(T)$ represents an integration of $\sigma_{1}$ over all frequencies, $\Sigma(T)$ is confined to an integration over our experimental bandwidth.  The difference between $\Sigma(T)$ and $\Sigma_{qp}(T)$ is plotted as triangles.  It is apparent that $\Sigma_{qp}(T)$ is less than the total spectral weight observed.  Notice that at low temperatures the difference, $\Sigma(T)-\Sigma_{qp}(T)$, is proportional to $\rho_{s}(T)$, shown here multiplied by 0.18.  The spectral weight in excess of $\Sigma_{qp}$ is proportional to $\rho_{s}$ only if we set $1/\tau_{qp}(T)\sim T$.  If we choose $1/\tau_{qp}(T)\sim T^{\beta}$ with any $\beta$ other than 1, the unaccounted for spectral weight does not have such a reasonable and recognizable $T$ dependence.  
  
The proportionality of the residual spectral weight and $\rho_{s}(T)$, suggests that the excess conductivity arises from fluctuations of the condensate order parameter.  The fluctuations are not thermally generated because their spectral weight increases as $T\rightarrow 0$.  Although the temperature dependence is consistent with intrinsic quantum fluctuations, the associated conductivity is expected to be of order $\sigma_{Q}$ or $\sim 1.6\times10^{5}(\Omega^{-1}-m^{-1})$\cite{Sachdev}, which is far too small to explain the data.  However, the fluctuation conductivity can be much larger in the presence of static (or quasistatic) spatial variation in the superfluid density.  Two examples have been discussed recently in the literature.  The first is a one-dimensional periodic modulation of the phase-stiffness parameter\cite{VanDerMarel}.  The second example is a "granular" superconductor in which the Josephson coupling between grains is randomly distributed about its mean value\cite{Barabash}.  Both of these models predict a contribution to $\sigma_{1}$ above zero frequency which is not present if the phase stiffness is homogeneous throughout the material.  Moreover, the spectral weight in this new contribution to $\sigma_{1}$ is proportional to the mean value of the phase stiffness, or superfluid density, just as is observed in our experiments.  In both cases the proportionality constant is $(\Delta\rho_{s})^{2}/\bar{\rho_{s}}^{2}$, the fractional mean square variation in $\rho_{s}$.

Motivated by these models, we attempt a decomposition of $\sigma_{1}$ into QP and collective mode (CM) contributions, $\sigma_{1}\equiv\sigma_{qp}+\sigma_{cm}$.  The difference between $\sigma_{1}$ and $\sigma_{qp}$ shows the collective mode contribution to be a low frequency peak contained, predominately, within our frequency range.  We therefore test this decomposition with $\sigma_{cm}$ chosen to be a Lorenztian centered at $\nu=0$.  This introduces two collective mode parameters: width, $\Gamma_{cm}(T)$, and spectral weight, $\rho_{cm}(T)$.  As is suggested by Fig. 2 we set the spectral weight of the collective mode, $\rho_{cm}(T)$, equal to a fixed fraction, $\kappa$, of $\rho_{s}(T)$.  For the normal fluid spectral weight we choose $\rho_{n}=\alpha T$ as before, and we vary $\Gamma_{cm}$ and $\tau_{qp}$ in order to find the best fit to the data.  The best fit to $1/\tau_{qp}(T)$ is shown in the upper right panel of Fig. 3.  (The dashed line indicates the range where the large thermal fluctuation conductivity dominates the QP contribution, making it difficult to extract $\tau_{qp}$). The width of the collective mode is found to be $T$ independent within the noise level, with a value 0.24 THz, and $\kappa = 0.30$.  
\begin{figure}[h]
     \includegraphics[width=3.25in]{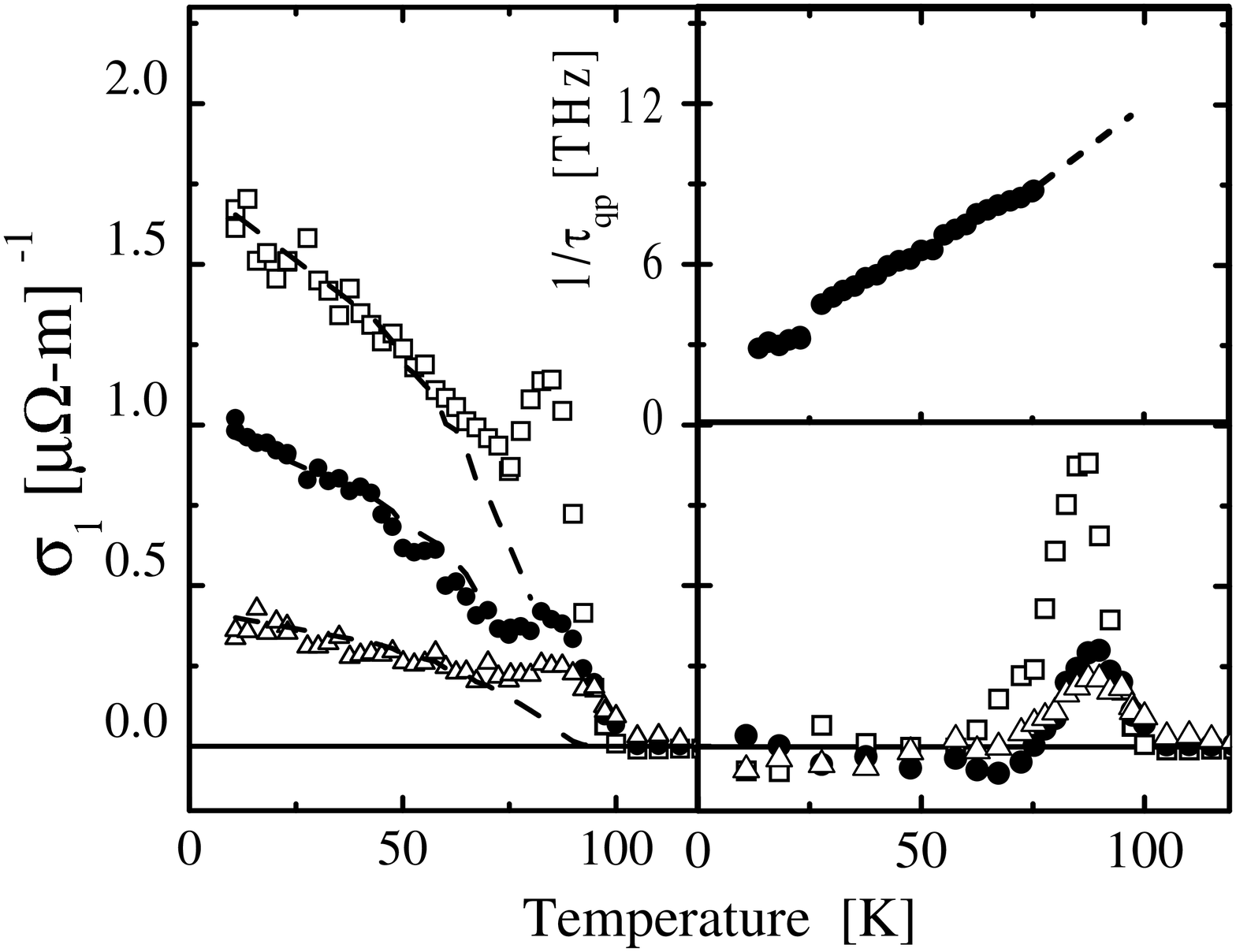}
\caption{Left panel: $\sigma_{1}(T)-\sigma_{qp}(T)$ plotted at 0.2, 0.36 and 0.64 THz as squares, circles and triangles, respectively.  The lines show $\sigma_{cm}(T)$ from our model.  Lower right panel: The difference between the data and the fit plotted on the same scale as the left panel, and ascribed to thermal phase fluctuations.  Upper right panel: $1/\tau_{qp}(T)$ for the QPs, due to this same model, plotted versus $T$ with the $T$ scale given by the panel below.}
\label{fig:Third}
\end{figure}

The left panel of Fig. 3 compares $\sigma_{1}(T)-\sigma_{qp}(T)$ with $\sigma_{cm}(T)$, obtained using the parameter values given above.  At each frequency $\sigma_{1}(T)-\sigma_{qp}(T)$ is plotted as symbols and $\sigma_{cm}(T)$ as a line.  The difference between the total conductivity and that due to QPs is obviously well described by the collective mode.  What deviation there is between the model and the data is shown in the bottom right panel of Fig. 3.  We see that difference corresponds to the phase fluctuations near $T_{c}$\cite{Nature}.    

At this point we comment on the collective mode conductivity spectrum.  In both models described previously\cite{VanDerMarel,Barabash}, the frequency of the collective mode is related to the screened plasma frequency of the condensate, or $\hbar\omega_{p}=\sqrt{\rho_{s}e^{2}/\epsilon d}$. With $\epsilon\sim 10$, this yields $\hbar\omega_{p}\sim 1400$ K, much higher than the energies probed in our experiment.  However, the plasma mode can be strongly perturbed by thermal QPs.  For example, the damping rate of the plasmon, $\Gamma$, is given by $\hbar\Gamma=\rho_{s}\sigma_{Q}/\sigma_{qp}$.  For the values of $\sigma_{qp}$ seen in our experiment, this is of order 100 K, indicating that the plasma mode is highly overdamped.  The corresponding conductivity will be centered at $\nu=0$ and its width will depend on the damping and screening effects of the QP background.  

One final observation concerning our data is that causality dictates that the collective mode feature in $\sigma_{1}(\nu)$ must also be clearly manifest in $\sigma_{2}(\nu)$.  Since both $\sigma_{1}$ and $\sigma_{2}$ are measured directly and independently by our coherent spectroscopy this is an important check of our model's description of the conductivity.  We find the total $\sigma_{2}(\nu)$ to be a sum of contributions from the QPs, the condensate and the collective mode.  $\sigma_{2}(\nu)$, therefore, is not described by a two-fluid model but rather by the model set forth above.     

In conclusion, we have shown that the dissipative conductivity of BSCCO in the superconducting state is not due solely to a normal fluid of QPs.  There is an additional contribution whose spectral weight increases with decreasing temperature.  We describe this contribution as a Lorentzian whose width is $T$ independent and whose spectral weight is a constant fraction of $\rho_{s}(T)$.  This additional dissipation could result from phase fluctuations in the presence of static or quasistatic spatial variations of the local superfluid density\cite{VanDerMarel,Barabash}.  From the spectral weight of this dissipation we infer a fractional mean square variation in $\rho_s$ of $\approx$ 0.30.  Adding this contribution to that of QPs with $1/\tau_{qp}\sim T$ successfully describes the conductivity over the entire experimental range of frequency.  Thus the transport lifetime in the superconducting state increases with decreasing $T$ far more slowly in BSCCO than in YBCO.  The difference in lifetime may arise from the presence of same inhomogeneities which generate the collective mode dissipation. 

We thank D.H. Lee, A. Maeda and D. Stroud for helpful discussions.  This work was supported by NSF Grant No. 9870258, DOE Contract No. DE-AC03-76SF00098, and ONR Contract No. N00014-94-C-001.

\end{document}